\documentclass[aps,prb,reprint,showpacs,superscriptaddress]{revtex4-1}

\usepackage{color}
\usepackage{graphicx}
\usepackage{amsmath}
\usepackage{amssymb}
\usepackage{dsfont}

\begin{document}

\title{Heat due to system-reservoir correlations in thermal equilibrium}
\date{\today}

\author{Joachim Ankerhold}
\email{joachim.ankerhold@uni-ulm.de}
\affiliation{Institut f\"ur Theoretische Physik, Universit\"at Ulm, Albert-Einstein-Allee 11, 89069 Ulm, Germany}

\author{Jukka P. Pekola}
\affiliation{ Low Temperature Laboratory (OVLL), Aalto University, School of Science, P.O. Box 13500, 00076 Aalto, Finland}

\begin{abstract}
The heat flow between a quantum system and its reservoir is analyzed when initially both are in a separable thermal state and asymptotically approach a correlated equilibrium. General findings are illustrated for specific systems and various classes of non-Markovian reservoirs relevant for solid state realizations. System-bath correlations are shown to be substantial at low temperatures even in the weak coupling regime. As a consequence, predictions of work and heat for actual experiments obtained within conventional perturbative approaches may often be questionable. Correlations induce characteristic imprints in heat capacities which opens a proposal to measure them in solid state devices.
\end{abstract}


\maketitle

\section{Introduction}

Recently, the subjects of heat and work and their distributions in the quantum regime have received considerable attention \cite{esposito:2009,campisi:2011} with measurement protocols proposed and implemented for solid state devices \cite{averin:2011,saira:2012, campisi:2013}. Classical thermodynamics considers a system of interest immersed in a much larger heat bath \cite{landau:2005}, where the latter one ensures thermal equilibration of the system, but  can be ignored otherwise. This weak coupling assumption works extremely accurately for basically all physical realizations on a macroscopic level. For quantum systems in contact with a heat bath realized e.g.\ through mesoscopic circuits, the situation is more intricate though. The non-locality of quantum mechanical wave functions induces system-reservoir correlations and even entanglement which may have profound impact also on thermodynamical properties \cite{weiss:2008}. Common wisdom is that at least in the weak coupling regime the classical-like setting of a thermal state factorizing in system and reservoir density operators, respectively, applies. Perturbative formulations for the reduced quantum dynamics of the system such as e.g.\ conventional master or Lindblad equations \cite{breuer:2007} rely on this separability.

Quantum mechanical system-environment correlations have been discussed in the past, mainly with respect to proper initial preparations for non-equilibrium dynamics, see e.g.\ Refs.~\cite{tannor2000,weiss:2008}. Much less attention has been paid to this subject in the context of quantum  thermodynamics \cite{lutz2009,lutz2011,lutz2011b,thingna2012}, where systems are kept close to thermal equilibrium and where the focus lies on quantities such as heat \cite{hanggi:2008} and work \cite{campisi:2011}.
   In many theoretical studies the simplified situation of (at least initially) factorizing thermal equilibria is taken for granted \cite{correra:2014,rossnagel:2014}. This way, predictions for work, work distributions, and heat due to the presence of weak classical driving sources have been obtained based on conventional master equations or related approaches \cite{averin:2011,pekola:2013,silaev:2013,suomela:2014}. However, in actual measurements, specifically in solid state structures, quantum correlations between system and reservoir may be of relevance not only far from but also close to and in thermal equilibrium.

The goal of this paper is to contribute to this latter topic.  For that purpose, we consider the situation, where an initially separable thermal state asymptotically approaches a correlated equilibrium and identify the exchanged energy to establish proper system-bath correlations as heat.  While an instantaneous switching-on of system-bath interactions is typically difficult to realize experimentally, it allows us, in a first step, to quantify the impact of these correlations for various types of reservoirs. We derive general expressions and discuss specific results for systems of possible experimental interest \cite{huber:2008,averin:2011,saira:2012, campisi:2013,rossnagel:2014}, namely, two level systems and harmonic modes. It turns out that even in the weak coupling regime, this heat flow is substantial at low temperatures and may become comparable to typical predictions for the work based on
conventional weak coupling approaches \cite{averin:2011,pekola:2013,silaev:2013,suomela:2014}. It further depends sensitively on non-Markovian features of the reservoir such that the commonly made simplification of a strictly ohmic environment \cite{breuer:2007} is always unphysical.

Hence, the so-defined heat is a profound measure for aspects of quantum thermodynamics beyond descriptions accounting merely for energy level quantization of otherwise separable systems and reservoirs \cite{deffner:2008}. Thus, in a second step, we show that the heat capacity of an embedded system encodes this information in form of e.g.\ a characteristic temperature dependence. It is proposed to measure it by manipulating and monitoring the reservoir in a solid state circuitry by means of advanced thermometry at cryogenic temperatures.

\section{General results and perturbation theory}

We consider a system and its surrounding in a standard setting with $H=H_S+H_R+H_C$. Since neither the system part $H_S$ nor the reservoir part $H_R$ do commute with the coupling $H_C$, the canonical thermal operator of the full compound
 \begin{equation}\label{fullequi}
W_\beta = \frac{{\rm e}^{-\beta H}}{Z}\,
\end{equation}
with corresponding partition function $Z$ describes system-bath correlations. The nature of these correlations has been an issue of intense research recently, see e.g.\ \cite{popescu2006,hur:2007,anders2008,lutz2009,kast:2013}, with the general conclusion that in many systems for sufficiently low temperatures they are non-classical and related to entanglement. This is not the case for a so-called factorized thermal state \cite{breuer:2007}
\begin{equation}\label{facequi}
W_f= \frac{{\rm e}^{-\beta H_S}}{Z_S}\otimes \frac{{\rm e}^{-\beta H_R}}{Z_R}\,
\end{equation}
where $Z_S$ and $Z_R$ denote the partition functions of isolated system and isolated bath, respectively.

We now think of a set-up, where one initially starts with a factorized state $W_f$  and then monitors the asymptotic state when the system has fully equilibrated to $W_\beta$. Due to the non-equilibrium initial preparation, energy will be exchanged between system and bath such as to establish proper equilibrium correlations between them.  Following the first law of thermodynamics in absence of external driving, this shift in system energy
\begin{eqnarray}
Q_{\rm corr}  &=& \langle H_S(t\to \infty)\rangle -\langle H_S(0)\rangle\nonumber\\
&=& \langle H_S\rangle_\beta -\langle H_S\rangle_f\, .
\end{eqnarray}
can be interpreted as heat due to system-bath correlations. Here, $\langle \cdot\rangle_{\beta/f}$ are expectation values taken according to the distributions (\ref{fullequi}) and (\ref{facequi}), respectively. Note that in contrast to the situation considered in \cite{lutz2011}, here, system-bath correlations do not appear due to an external (adiabatic) switching-on of the system-bath interaction but solely due to the intrinsic dynamics according to the full Hamiltonian $H$. Accordingly, the reduced density operator of the system alone $\rho(t)={\rm Tr}_R\{W(t)\}$ evolves from a Gibbs state into a thermal state which in general is not of Gibbs form.

A formal expression can be found based on
the relation between the respective partition functions and energy expectation values, i.e.\
\begin{equation}\label{qexacta}
Q_{\rm corr} =-\frac{\partial}{\partial\beta}{\rm ln}(Z/Z_S) -\langle H_R+H_C\rangle_\beta\, .
\end{equation}
While for specific systems such as harmonic oscillators, this expression can be evaluated exactly, in general,
a perturbative treatment must be applied. In the sequel, we focus on the weak coupling regime and calculate the heat exchange in lowest non-vanishing order in $H_C$ (canonical perturbation theory \cite{tannor2000}). This is conveniently done by iterating the operator identity
\begin{equation}
{\rm e}^{-\beta H} =  {\rm e}^{-\beta H_0} \left( 1- \int_0^\beta d\lambda\, {\rm e}^{\lambda H_0}\, H_C\, {\rm e}^{-\lambda H}\right)
\end{equation}
with $H_0=H_S+H_R$. Since in equilibrium, the bath is assumed to exert no net force, i.e.\ $\langle H_C\rangle_f=0$, one needs to iterate at least up to the second order in $H_C$. This yields
\begin{equation}
Z\approx Z_S Z_R \left[ 1+ \frac{1}{\hbar^2}\int_0^{\hbar\beta} d\lambda \int_0^\lambda d\tau\,  \langle H_C(-i\tau) H_C\rangle_f\right]\,
\end{equation}
with Heisenberg operators taken with respect to $H_0$. When combined with a similar result for the not-normalized energy expectation values one finds
\begin{eqnarray}
Q_{\rm corr} &\approx &  \frac{1}{\hbar^2}\int_0^{\hbar\beta} d\lambda \int_0^\lambda d\tau\,  \Big[\langle H_S\, H_C(-i\tau) H_C\rangle_f \nonumber\\
&&- \langle H_S\rangle_S\, \langle H_C(-i\tau) H_C\rangle_f \Big]\, .
\end{eqnarray}
This result applies to arbitrary system operators when $H_S$ is replaced accordingly. Here and in the sequel, $\langle\cdot\rangle_{R, S}$ denote expectation values based on the individual canonical operators of reservoir and system, respectively.

To proceed, we consider the generic situation of a bath with Gaussian noise properties bilinearly coupled to a system via $H_C= q {\cal E}$ with ${\cal E}$ being a collective bath mode and $q$ a dimensionless system operator. The reservoir is then completely determined by the equilibrium correlation $L(t)=\langle {\cal E}(t) {\cal E}\rangle_R $ so that
\begin{eqnarray}\label{qcorr}
Q_{\rm corr}&=& \frac{1}{\hbar^2}\int_0^{\hbar\beta} d\lambda \int_0^\lambda d\tau\,  L(-i\tau) \Big[\langle H_S\, q(-i\tau) q\rangle_S\nonumber\\
&& - \langle H_S\rangle_S\, \langle q(-i\tau) q\rangle_S \Big]\, .
\end{eqnarray}
In imaginary time the bath correlation takes the form \cite{weiss:2008}
\begin{equation}\label{kernel}
L(-i\tau) = \mu\, :\delta(\tau): - k(\tau)\, ,
\end{equation}
 where $:\delta():$ denotes a periodically continued $\delta$-function beyond the interval $\hbar\beta$ and
$\mu = (2/\pi) \int_0^\infty d\omega I(\omega)/\omega$
with the spectral distribution of bath modes $I(\omega)$. Further, the kernel
\begin{equation}\label{imkernel}
k(\tau)=\frac{1}{\beta} \sum_{n=-\infty}^\infty \zeta_n\, {\rm e}^{i\nu_n \tau}\,
\end{equation}
is periodic in $\hbar\beta$ and
\begin{equation}
\zeta_n=|\nu_n| \hat{\gamma}(|\nu_n|)
 \end{equation}
 contains the Matsubara frequencies $\nu_n=2 \pi n/\hbar \beta$ and the Laplace transform of the classical damping kernel
\begin{equation}
\hat{\gamma}(z)=\frac{2}{\pi}\int_0^\infty d\omega \frac{I(\omega)}{\omega} \frac{z}{z^2+\omega^2}\, .
\end{equation}
Note that this perturbative treatment is valid as long as $Q_{\rm corr}$ is sufficiently smaller than typical bare level spacings of the system of interest.

\section{Two-level system}
According to common experimental realizations in superconducting circuits, we first discuss a two level system \cite{shnirmann2001,campisi:2013}
\begin{equation}\label{tls}
H_S=-\frac{\hbar\Delta}{2} \, \sigma_x
\end{equation}
with coupling operator $q=\sigma_z$. Performing the time integrations in (\ref{qcorr}) with (\ref{imkernel})
yields with the bare result
\begin{equation}
\langle H_S\rangle_S = -\frac{\hbar\Delta}{2} {\rm tanh}(\theta/2)
\end{equation}
the expression
\begin{equation}\label{qexact}
Q_{\rm corr}= \frac{\hbar\Delta^2}{2{\rm cosh}^2(\theta/2)} \sum_{n\geq 1} \frac{\zeta_n}{\Delta^2+\nu_n^2}\, \left[1-\frac{\Delta^2-\nu_n^2}{\Delta^2+\nu_n^2} \frac{{\rm sinh}(\theta)}{\theta} \right]
\end{equation}
with the dimensionless inverse temperature $\theta=\Delta\hbar\beta$. Reservoir properties only appear in $\zeta_n$ so that this result allows to analyze $Q_{\rm corr}$ for various spectral distributions.
\begin{figure}
\centering
\includegraphics[width=1.02\columnwidth]{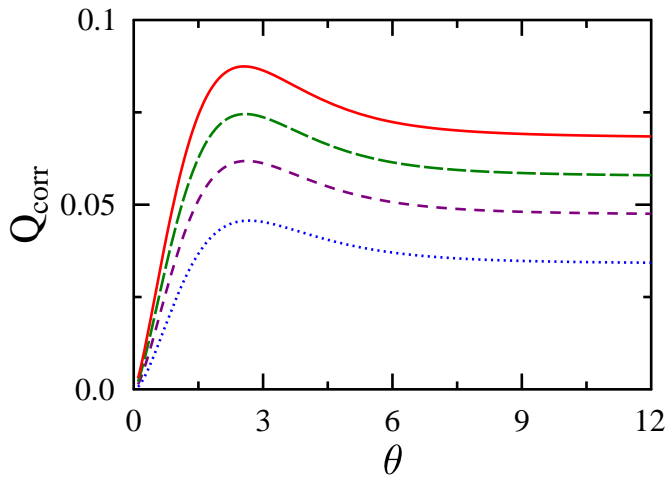}
\caption{\label{fig1} Heat $Q_{\rm corr}$  from (\ref{qexact}) exchanged between system and an ohmic-type reservoir to establish equilibrium correlations from an initially factorized state vs.\ inverse temperature $\theta=\Delta\hbar\beta$ for various cut-off frequencies $\omega_c/\Delta=$ 20 (blue, dotted), 50 (purple, short-dashed), 100 (green, long-dashed), 200 (red, solid) and coupling constant $\eta=0.1$. Heat is scaled with $\hbar\Delta$.}
\end{figure}
\begin{figure}
\centering
\includegraphics[width=1.02\columnwidth]{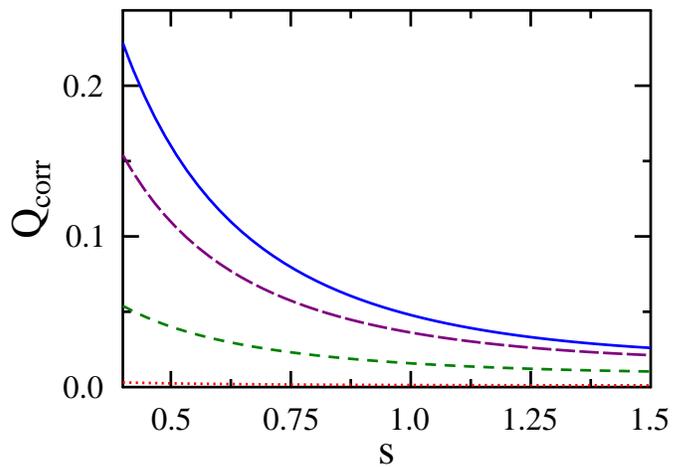}
\caption{\label{fig2} Heat $Q_{\rm corr}$  exchanged between system and sub-ohmic reservoirs with spectral exponents $s$ [see (\ref{specs})]  to establish equilibrium correlations from an initially factorized state at various inverse temperatures $\Delta\hbar\beta=$ 0.1  (red, dotted), 0.5 (green, short-dashed), 1 (purple, long-dashed), 10 (blue, solid). Heat is scaled with $\hbar\Delta$ and $\eta=0.1$, $\omega_c/\Delta= 50$.}
\end{figure}
We start with a Drude bath
\begin{equation}\label{drude}
I(\omega)= \eta \frac{\omega \omega_c^2}{\omega_c^2+\omega^2}\,
\end{equation}
with coupling parameter $\eta$ and Drude frequency $\omega_c$. This then implies $\hat{\gamma}(z)=\eta\omega_c/(\omega_c+z)$. Numerical results are shown in Fig.~\ref{fig1}.  At high temperatures $\theta\ll 1$, the heat grows quadratically
\begin{equation}\label{qhigh}
Q_{\rm corr, high}\approx  \frac{\eta \hbar\omega_c}{4\pi^2}\, \theta^2\,
\end{equation}
so that in this regime, as expected, a factorizing initial state is an accurate approximation. In contrast, at moderate and low temperatures the exchanged heat is quite substantial even for weak coupling and it tends to saturate at very low temperatures $\theta\gg 1$. For $\omega_c\gg \Delta$ the leading contributions read
\begin{equation}\label{qinfi1}
Q_{{\rm corr, low}} \approx \frac{\eta\hbar\Delta}{2\pi}\, {\rm ln}(\omega_c/\Delta) +\frac{\eta\hbar\Delta}{4 \omega_c}-\frac{\eta\hbar\Delta}{\theta^2}\frac{\pi}{6}\, .
\end{equation}
Here, the logarithmic dependence on the Drude frequency reflects the impact of zero-point fluctuations of the reservoir (Lamb-shift) which have recently been measured in a circuit quantum electrodynamical set-up \cite{Fragner:2008}. It originates from the second term in brackets in (\ref{qexact}) while the first one dominating the classical and moderate quantum regime is exponentially suppressed. The changeover to the deep quantum domain roughly occurs when $\cosh(\theta/2)\approx 1$, i.e.\ $\theta\approx 2$, and according to Fig.~\ref{fig1} is related to a maximum in $Q_{\rm corr}$. Note that in a strict ohmic limit $\omega_c\to \infty$ the exchanged heat diverges and non-Markovian properties must {\em always} be taken into account. Treatments which not only assume factorized initial states but also purely Ohmic reservoirs to predict heat and work distributions for open systems may thus be of only limited value at low temperatures. Further, the result (\ref{qinfi1}) reveals that the expression (\ref{qexact}) applies even at zero temperature as long as $\eta {\rm ln}(\omega_c/\Delta)\ll 1$.

Let us now turn to more general types of spectral distributions
\begin{equation}\label{specs}
I(\omega) =  \eta \frac{\omega^s\, \omega_c^{3-s}}{\omega_c^2+\omega^2}
\end{equation}
characterized by a spectral exponent $0\leq s<2$. For $s<1$ (sub-ohmic noise) this distribution describes a class of reservoirs which appears in quantum optical \cite{porras:2008}  and mesoscopic set-ups \cite{tong:2006} and for $s\ll 1$ mimics $1/f$-noise, a significant noise distribution in the low temperature regime of solid state devices \cite{paladino2014}. The case $s>1$ is known as super-ohmic decoherence. It has recently been shown that in the sub-ohmic regime at sufficiently low temperatures, system and bath are strongly correlated due to entanglement \cite{hur:2007,kast:2013}. As illustrated in Fig.~\ref{fig2}, this has direct consequences for the heat generated from an initially factorized state. In fact, in the zero temperature limit one arrives  in leading order in $\omega_c/\Delta$ at
\begin{equation}
Q_{{\rm corr}, 0}(s) \approx \frac{\eta\hbar\Delta}{2\sin(\pi s)} \left[(\omega_c/\Delta)^{1-s}-\sin(\pi s/2)\right]\, .
\end{equation}
Here, the limit $s\to 1$ must be taken with care to regain (\ref{qinfi1}). As seen in Fig.~\ref{fig2}, even reservoirs with moderate sub-ohmic characteristics display an enhanced heat production at low temperatures which is much less pronounced in the super-ohmic case and basically absent at high temperatures. In the regime $s\ll 1$, relevant for cryogenic solid state experiments, heat is mainly determined by reservoir energy scales, i.e.,
$Q_{{\rm corr}, 0}(s\ll 1) \approx (\eta\hbar\omega_c)/(2\pi s)$.

\section{Exact results: Harmonic oscillator}

Exact results valid also for strong system-bath coupling can be obtained for harmonic systems. This not only allows to inspect the validity of weak coupling predictions, but may also be of experimental relevance \cite{huber:2008,rossnagel:2014}, in solid state systems e.g.\ for circuits including Josephson junctions such as superconducting quantum interference devices (SQUID). This way, we consider
\begin{equation}\label{harmH}
H_S= \frac{p^2}{2m}+\frac{m\omega_0^2}{2} q^2
\end{equation}
which yields under a factorized equilibrium
\begin{equation}\label{harm1}
\langle H_S\rangle_S =\frac{\hbar\omega_0}{2}\, {\rm coth}(\omega_0\hbar\beta/2)\, .
\end{equation}
Instead, in correlated equilibrium one has \cite{weiss:2008}
\begin{equation}
\langle H_S\rangle_\beta = \frac{1}{\beta}+\frac{1}{\beta} \sum_{n\geq 1} \frac{2\omega_0^2+\zeta_n}{\nu_n^2+\omega_0^2+\zeta_n}\, ,
\end{equation}
with the second term describing quantum mechanical fluctuations.
Hence, one finds for the exchanged heat (\ref{qexact}) the exact expression
\begin{equation}\label{harmqex}
Q_{\rm corr, ex}= \frac{1}{\beta}\sum_{n\geq 1}\frac{\zeta_n}{\nu_n^2+\omega_0^2+\zeta_n}\, \frac{\nu_n^2-\omega_0^2}{\nu_n^2+\omega_0^2}\, .
\end{equation}
\begin{figure}
\centering
\includegraphics[width=1.02\columnwidth]{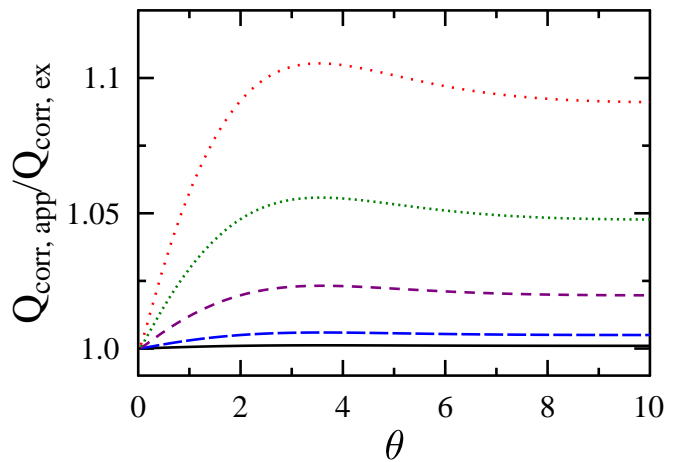}
\caption{\label{fig3} Approximate value of the exchanged heat for weak coupling $Q_{\rm corr, app}$ for a harmonic oscillator normalized to the exact value $Q_{\rm corr, app}$ vs. inverse temperature $\theta$ in a Drude bath ($\omega_c/\omega_0=50$).  Depicted are results for various coupling strengths $\eta = 0.01$ (black, solid), 0.05 (blue, long-dashed), 0.2 (purple, short-dashed), 0.5 (green, dotted-closely spaced), and 1 (red, dotted).}
\end{figure}
In the regime of weak coupling, this reduces according to (\ref{qcorr}) to
\begin{equation}\label{harmapprox}
Q_{\rm corr, app}\approx \frac{1}{\beta}\sum_{n\geq 1} \frac{\zeta_n}{\nu_n^2+\omega_0^2}\, \frac{\nu_n^2-\omega_0^2}{\nu_n^2+\omega_0^2}\,
\end{equation}
which can further be approximated for a Drude model (\ref{drude}).  At high temperatures $\theta\ll 1$ one gains
\begin{equation}\label{harmhigh}
Q_{\rm corr, high}\approx \frac{\eta \hbar\omega_c}{4\pi^2} \, \theta\,
\end{equation}
 with a linear rise instead of a quadratic one in (\ref{qhigh}) due to an unbounded energy spectrum. The factorizing assumption seems thus to be better justified for two-level systems than for systems with an infinite number of accessible states. In the zero temperature regime and for weak coupling, the harmonic oscillator reduces to a two level system which in turn may verify the validity of the result (\ref{qinfi1}). Indeed, the exchanged heat from (\ref{harmapprox}) coincides with the result (\ref{qinfi1}) with $\Delta$ replaced by $\omega_0$.

To analyze the accuracy of the weak coupling treatment for finite temperatures, we depict in Fig.~\ref{fig3} the ratio of the approximated result (\ref{harmapprox}) to exact one (\ref{harmqex}). Notably, the perturbative treatment works fairly accurately also for somewhat stronger dissipation and over the full temperature range. Its predictions tend to deviate substantially from the exact values, however, in the overdamped regime $\eta>1$. In this regime, the reservoir induced level broadening $\eta\hbar\omega_0$ even exceeds the bare level spacing so that for very strong dissipation and at low temperatures  $\eta\gg 1$, $\omega_c\hbar\beta\gg \omega_0\hbar\beta\gg 1$ one finds from (\ref{harmqex})
\begin{equation}
Q_{\rm corr, over}\approx \frac{\eta \hbar\omega_0}{2\pi}\, {\rm ln}(\eta\omega_c/\omega_0)\, .
\end{equation}
Note that in this latter regime, it is known that entanglement correlates system and bath \cite{anders2008,lutz2009}

\section{Heat capacity}

Heat capacity is a central experimental quantity in bulk systems. It has thus been proposed  as measure to access information about entanglement \cite{wiesniak:2008}. Recently, this has indeed been demonstrated for large spin ensembles \cite{singh:2013}.

Here, we follow a somewhat more ambitious route and consider the heat capacity of a single quantum degree of freedom embedded in a thermal bath. We show to what extent system-bath correlations induce deviations from bare predictions. Accordingly, we consider heat capacities $C_\beta$ in full thermal equilibrium and compare them against those of a bare system thermal state $C_S$. Both follow from the temperature dependence of the respective system energy expectation values (internal energy), the difference of which $Q_{\rm corr}$ has been calculated above.  Classically or for  bare systems, heat capacities can be obtained from the temperature dependence of either internal energies or partition functions.  In full thermal equilibrium, however, only the energy-based definition leads to physically meaningful predictions \cite{hanggi:2008}. We thus obtain
\begin{equation}
\frac{C_\beta}{C_S}= 1+ \frac{\partial Q_{\rm corr}(T)}{C_S\, \partial T}\, .
\end{equation}

For the two level system (\ref{tls}), the bare text-book expression \cite{landau:2005} reads
\begin{equation}\label{cbare}
\frac{C_S}{k_{\rm B}} = \frac{\theta^2}{4 {\rm cosh}^2(\theta/2)}\, .
\end{equation}
\begin{figure}
\centering
\includegraphics[width=1.02\columnwidth]{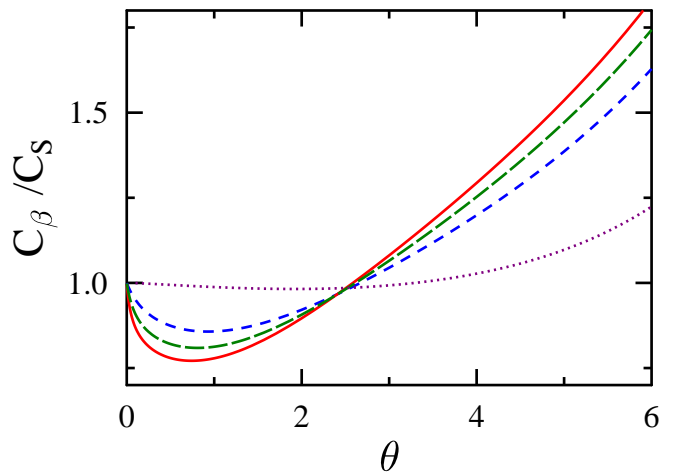}
\caption{\label{fig4} Ratio of the heat capacities $C_\beta/C_S$ of correlated and bare thermal state vs.\ inverse temperature for a two level system and various bath cut-off frequencies $\omega_c/\Delta=$ 20 (blue, short-dashed), 50 (green, long-dashed), 100 (red, solid). For comparison also results for a harmonic system are depicted (purple, dotted) for $\omega_c/\omega_0=50$. Coupling strength is $\eta=0.1$.}
\end{figure}
\begin{figure}
\centering
\includegraphics[width=1.02\columnwidth]{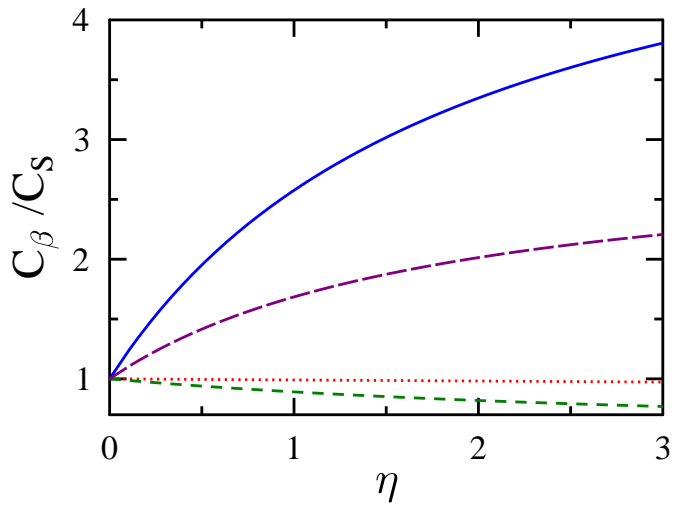}
\caption{\label{fig5} Ratio  $C_\beta/C_S$ of correlated and bare thermal state of a harmonic oscillator vs.\ coupling strength for various inverse temperatures $\omega_0\hbar\beta=$ 0.1 (red, dotted), 1 (green, short-dashed), 4 (purple, long-dashed), 6 (blue, solid).}
\end{figure}
This result can now be compared with the full heat capacity stemming from (\ref{qexact}), see Fig.~\ref{fig4}. In particular, based (\ref{qhigh}), in  the high temperature limit both basically coincide
\begin{equation}
\frac{ C_\beta}{C_S} \approx 1-\frac{2\eta \omega_c}{ \pi^2} \, \theta\, .
\end{equation}
However, at very low temperatures,  according to (\ref{qinfi1}),
\begin{equation}\label{heatlow}
\frac{ C_\beta}{C_S} \approx \frac{\eta\omega_c \pi}{3\theta^3} \, {\rm e}^{\theta}\, .
\end{equation}
 One observes a strong enhancement due to the fact that the full $C_\beta$ decays only algebraically in contrast to (\ref{cbare}).
For the full heat capacity the changeover between the classical/moderate quantum to the deep quantum regime is related to substantial suppression and enhancement, respectively, compared to the bare prediction. This feature is more pronounced for large Drude frequencies.

For harmonic systems (\ref{harmH}), the well-known bare result
\begin{equation}
\frac{C_S}{k_{\rm B}} =  \frac{\theta^2}{4\sinh^2(\theta/2)}
\end{equation}
leads for high temperatures together with (\ref{harmhigh}) to
\begin{equation}
\frac{C_\beta}{C_S}\approx 1-\frac{ \eta\hbar\omega_c}{4 \pi^2}\, \theta^2\, .
\end{equation}
For  low temperatures and weak coupling one regains the expression (\ref{heatlow}). Exact data can easily be obtained from (\ref{harmqex}).
Figure~\ref{fig4} reveals an only weak temperature dependence of $C_\beta/C_S$ in the range shown and the overall behavior is basically insensitive to $\omega_c$ for $\omega_c/\omega_0\gg 1$. The strong coupling domain, where perturbative results are not applicable,  is addressed in Fig.~\ref{fig5}. Interestingly, larger coupling has not always the tendency to increase the heat capacity compared to the bare one even at moderately low temperatures. One obtains a substantial enhancement only for very low temperatures. Notably, this occurs in the regime $\eta>1$ and $\eta\omega_0\hbar\beta\gg 1$, also known as the quantum Smoluchowski limit \cite{ankerhold:2001}, where the level broadening due to friction exceeds both the bare level spacing and the thermal energy scale so that quantum dynamics tends to become more classical, though,  with substantial quantum fluctuations.

\section{Experimental detection}

We now propose a scheme to retrieve information about the impact of system-bath quantum correlations by measuring the heat capacity of the embedded system. The latter one is taken as a solid state implementation of a two level system, e.g.\ in form of a biased SQUID coupled inductively to a resistor (reservoir). In these devices, at cryogenic temperatures, electronic degrees of freedom (Fermi gas) are very weakly coupled to the underlying phonon background with relaxation times on the order of $100\,\mu$s while electron-electron interaction leads to equilibration in the Fermi gas within a few nanoseconds. It is this latter heat bath which dominantly interacts with the SQUID. By means of rf-thermometry \cite{cryo1,cryo2,cryo3}, it is now possible to monitor the actual temperature of the Fermi gas of a mesoscopic metallic island on sub-$\mu$s time scales which in turn allows to monitor its relaxation dynamics  after a heating pulse has been applied.

We imagine a situation, where both SQUID and its electronic environment are in thermal equilibrium at a temperature $T_1$. A short and weak heating pulse of duration $\tau_p$ is sent to the reservoir, where it leads within a few nanoseconds to a  temperature $T_2$ of the electron gas. The coupling between the system and this reservoir is assumed to be weak such the coupling rate $\Gamma$ depending on their correlations obeys $\tau_p\sim 1/\Gamma_{\rm ee}\ll 1/\Gamma\ll 1/\Gamma_{\rm ep}$ with $\Gamma_{\rm ee}$ ($\Gamma_{\rm ep}$) being the equilibration rate of the Fermi gas (Fermi gas and phonon bath). Accordingly, on a much shorter time scale than $1/\Gamma_{\rm ep}$ the electronic reservoir will transfer energy to the two-level system  and eventually equilibrate with it. This loss of energy of the electronic reservoir corresponds to a heat flow $Q_R= C_R \delta T$, where $C_R$ is the heat capacity of the bare reservoir and $\delta T$ is the temperature drop $T_2\to T_2-\delta T$ due to the reservoir-system equilibration. This heat flow balances the heat flow received by the system when it heats up from $T_1\to T_2-\delta T$, i.e., $C_\beta (T_2-\delta T-T_1)$ with $C_\beta$ being the heat capacity of the embedded system. Hence, one arrives at
\begin{equation}
\frac{C_\beta}{C_R}\approx \frac{\delta T}{T_2-T_1-\delta T}\, ,
\end{equation}
where the small portion of heat lost to the phonon-bath during the equilibration between system and reservoir has been neglected. It can be estimated to be on the order of $\Gamma_{\rm ep}/\Gamma\ll 1$. A specific advantage of this protocol is that the heat capacity $C_R$ (known for typical bulk materials) can be extracted {\em in situ} from the known energy carried in the initial heating pulse $E_{\rm in}$ via $C_R=E_{\rm in}/(T_2-T_1)$. Experimental data for varying $T_1$ and $T_2$ (but still small $T_2-T_1$) can then be compared with predictions for the bare system according to the results of the previous section.

\section{Conclusion}

The quantum correlations between a system and its reservoir are analyzed in terms of the heat exchange during the equilibration when
starting initially  with a separable thermal state. Specific results are discussed for systems for which recent theoretical predictions of work, work distributions and heat  have been made based on factorized thermal states. System-bath correlations substantially influence heat capacities in the low temperature regime which may open a way to detect them in solid state devices.

As we have shown,  system-bath correlations induce deviations in the weak coupling quantum regime when compared to predictions based on separable thermal states on the order of 10\% of the bare level splitting. When a weak external driving is exerted to the system, according to the first law of thermodynamics, part of its energy is deposited into the system (internal energy) and part of it is transferred to the bath in form of heat. Weak coupling approaches such as master equations obtain work and heat based on separable thermal equilibria \cite{pekola:2013,silaev:2013,suomela:2014}. For example, for a monochromatic pulse with frequency $\Omega$ and amplitude $\lambda_0$ in resonance with a two-level system, they are applicable as long as work and heat are on the order of $(\lambda_0/\hbar\Omega)^2\ll 1$ which basically matches the impact of system-bath correlations. For actual realizations at low temperatures, their predictions may thus be of limited reliability. System-bath correlations may be substantial not only in theoretical approaches to understand work and heat at the quantum level but also to analyze experimental data.

\acknowledgements
The work has been supported partially by the DFG through AN336/6-1, by the Academy of Finland
through its LTQ (Project No. 250280) CoE Grant, the AScI visiting professor program at Aalto University,
European Union FP7 Project INFERNOS (Grant Agreement No. 308850).

\bibliographystyle{apsrev4-1}
\bibliography{bibheat}

\end{document}